# A NOVEL FRAMEWORK FOR DISINFECTION ANALISYS IN DRINKING WATER NETWORKS

D. Laucelli[1], L. Vergine[2], G. Messa[3], O. Giustolisi[4]*

[1]Politecnico di Bari, via Orabona, 4, Bari, Italy, danielebiagio.laucelli@poliba.it
[2]Acquedotto Pugliese S.p.A., via Cognetti, 36, Bari, Italy, l.vergine1@phd.poliba.it
[3]Acquedotto Pugliese S.p.A., via Cognetti, 36, Bari, Italy, g.messa1@phd.poliba.it
[4]Politecnico di Bari, via Orabona, 4, Bari, Italy, orazio.giustolisi@poliba.it
*(corresponding author)*

## Abstract

Disinfection in drinking water networks is performed to ensure water safety and potability. However, disinfectants can react with organic compounds present in the water networks. The reaction of disinfectants with such compounds generates byproducts that could be dangerous for human health. The compounds are generally endogenous, i.e., adhered or/and released by the pipe wall, and not introduced from reservoirs because water utilities efficiently monitor and manage their water quality. The present effort, starting from the relevant scientific literature, proposes a novel framework which is based on analysing the disinfectant consumption using a second order kinetic model. The novel framework allows analysing the two endogenous mechanisms of compounds reactions: the first mechanism refers to compounds detached by the momentum of water flow and reacting into the bulk while transported by flow, the second one refers to local reaction at wall.

The use of a chemically based second order kinetic model, furthermore, allows for emphasizing the role of stoichiometry and reaction rate constants of reactants and byproducts as relevant parameters to be evaluated at laboratory scale for the specific water system. Thus, the proposed framework can effectively support water quality management using monitoring data together with hydraulic modelling. Two real water distribution systems, located in southern Italy and managed by Acquedotto Pugliese, were used as case studies: the small-sized network of Monteparano and the large-sized one of Bari.

# 1 Introduction

The disinfection is a relevant issue for drinking water networks security and for human health. Incorrect disinfection can cause excessive production of disinfection by-products (*dbp*s). The main *dbp*s include, among others, inorganic by-products, organic oxidation by-products, and halogenated organic by-products, which are potentially harmful for human health (Azara et al., 2010). Therefore, national regulations require monitoring the levels of organic compound, *dbp*s and disinfectant residues into the network (Mazhar et al., 2020). Over the years, these regulations have induced water utilities to develop operational protocols that guarantee the minimization of organic compounds, *dbp*s and the associated health risk (LeChevallier et al., 1996; Power and Nagy, 1999).

At European level, the novel EU Directive n. 2184 (EU Council, 2020) introduces obligations on the risk assessment and management for drinking water distribution networks (WDNs). Then, water utilities are committed to assess the contamination risk and manage disinfection, from the water source(s) to consumers. The novel EU Directive makes realistic the perspective of water quality management because it emphasizes the role and the responsibility of water utilities with respect to human health. Therefore, the continuous monitoring of residual disinfectant, organic compounds and *dbp*s formation into WDNs is a realistic future perspective (Qiu et al., 2020). To be prepared to the collection of water quality data, a novel framework for disinfection analysis is useful to transform those data into information for the effective management of disinfectant, organic compounds and *dbp*s.

Many researchers have been committed from more than two decades in studying disinfection in WDNs from different standpoints. Boccelli et al. (2003), Shang et al. (2008) and Qiu et al. (2020) for example, pointed out the need to go beyond the single-species water quality models, assuming that chlorine (i.e., the most used disinfectant) reacts with organic matter, which is a heterogeneous mixture of organic compounds, as well as with other inorganic substances that may be present in the water and at the pipes wall. Abhijith and Ostfeld (2022) defined a new framework to overcome some critical challenges related with the time-series behaviour of multiple species within the spatial boundaries of water distribution systems.

The Clark's mathematical formulation (Clark, 1998) of the second order kinetic model has been relevant to develop the proposed novel framework. Clark considered chlorine, organic matter and trihalomethanes as *dbp*s, overcoming the limits of the first order kinetic model. The novel framework based on the second order kinetic model allows the presentation of a new mechanism of endogenous reacting compounds, also.

The wall of aged pipes plays a crucial role in disinfectant decay, because of its (i) direct reaction at the wall and (ii) possible wall biofilm detachment due to water flow momentum (Bonadonna et al., 2008; Lehtola et al., 2006; Tsai, 2005). Therefore, these findings have been useful to develop the concept of two distinct endogenous mechanisms capturing the reason of the disinfectant decay. In fact, the novel framework allows analysing the compound detached from the wall biofilm by the momentum of water flow reacting into the bulk, while transported by flow, and the local reaction at wall biofilm.

This paper will demonstrate and discuss that the two different endogenous mechanisms, related to the pipe wall biofilm, generate two distinct patterns of disinfectant decay into the network. The previously reported chemical framework is integrated with advanced hydraulic analysis (Giustolisi et al., 2008; Giustolisi et al., 2023) and an effective Lagrangian scheme of transport and decay (Giustolisi et al., 2023). Note that the Lagrangian scheme, based on the accurate calculation of the velocity field using an unlimited number of pipe parcels (Giustolisi et al., 2023) , is important to exploit the novel proposed chemically based framework.

The paper is organized as follows: the next paragraph reports the relevant scientific literature on the disinfection process; the third paragraph develops and discusses the mathematical framework of second order kinetic model and the implementation in hydraulic modelling; the fourth paragraph introduces and discusses the two different endogenous mechanisms, due to the wall biofilm generating two distinct patterns of disinfectant decay into the network; the fifth paragraph briefly reports the hydraulic modelling approach and the relevant characteristics of the Lagrangian scheme used; the sixth paragraph demonstrates the two distinct patterns using the small-sized hydraulic system of Monteparano; the seventh paragraph discusses disinfectant management perspectives using the large-sized hydraulic system of Bari; finally, concluding remarks and perspective are presented.

## 2 Relevant scientific literature for the proposed framework

From a water quality modelling perspective, there are two significant physical decay mechanisms due to the biofilm of aged pipes: one in the bulk water and one at wall (Hallam et al., 2002). The bulk water species are chemical or biological compounds transported through the network by the water flow, while pipe wall species are compounds attached or embedded, fixed or partially fixed, that the flow condition can detach becoming bulk water species (Lehtola et al., 2006). Note that the name biofilm refers to compounds attached at the wall, including bacteria, metal ions from corrosion, etc. (Bonadonna et al., 2008).

The most used disinfectant in WDNs is chlorine (Vasconcelos et al., 1996). The advantages of chlorine disinfection include simplicity of application, low cost and effectiveness. Disinfectant consumption is influenced by several factors, mainly by the quantity of organic and inorganic reacting compounds, which are present in the bulk water and constitutes the wall biofilms (Lehtola et al., 2006). Water temperature, pH and contact time can also influence the decay mechanisms (Powell et al., 2000).

The reaction mechanisms of a disinfectant are complex and related to site-specific conditions and heterogeneous nature of reacting compounds (Boccelli et al., 2003). Bulk decay was initially analysed using the first order kinetic model (Hallam et al., 2003; Rossman et al., 1994; Haas and Karra, 1984). To the purpose of the discussion of the novel proposed framework, a simple and general formulation of first order kinetic model is here reported,

$$\frac{dC_{Cl}}{dt} = -k_{Cl}C_{Cl} \tag{1}$$

where $C_{Cl}$ is the chlorine concentration in [g/m$^3$] and $k_{Cl}$ is the first order reaction rate constant [d$^{-1}$]. Other proposed models are the pseudo second order model (Powell et al., 2000), the pseudo n order model (Haas and Karra, 1984; Powell et al., 2000) and the parallel first order model (Haas and Karra, 1984; Chang et al., 2006). Note that first order kinetic models assume that the concentration of reacting compounds is in excess with respect to chlorine and the decay rate depends only on the chlorine concentration. However, subsequent studies have shown that this type of models do not effectively represent water quality conditions in WDNs (Wang et al., 2019). Furthermore, the absence of information, see Eq. (1), about the reacting compounds and *dbp*s formation, makes the first order kinetic model ineffective for water quality management.

Wall decay was initially simply modelled using zero order or first order kinetic models, but subsequent studies have shown that biofilm activity can be modelled differently, and it plays a significant role in chlorine reaction at pipe wall (Hallam et al., 2002; Digiano and Zhang, 2005).

For these reasons, the second order kinetic model has been proposed for over two decades. Among these are the second order kinetic model for chlorine and organic compound decay and *dbp*s production proposed by Clark (1998), the two-reagent second order kinetic model (Fisher et al., 2012; Fisher et al., 2011; Jabari Kohpaei and Sathasivan, 2011) and the variable rate coefficient model (Jonkergouw et al., 2009). However, it is well known that the decay mechanism of any disinfectant is composed by a complex set of parallel and serial, fast and slow, reactions (Kastl et al., 1999) and, therefore, the second order kinetic model represents a synthetic model. The parameters of such second order kinetic model, being chemically based, can be calibrated using monitoring data for the specific WDN and could be transferred as priors to other WDNs (Fisher et al., 2017; Wang et al., 2019) , before a specific tuning.

Fisher et al. (2017) developed an advanced chlorine decay model that integrates biofilm activity, offering a more realistic depiction of chlorine behaviour in water distribution systems. Initially, chlorine is mainly consumed in the bulk water (large pipes, high chlorine), but further downstream, wall reactions dominate and eventually is limited by residual chlorine concentration.

Wang et al. (2019) investigated the bulk decay of chlorine in reclaimed water samples with different quality, proposing a series of indices to directly characterize and quantify the chlorine-reactive substances in WDN carrying reclaimed water through a second order decay model derived from the model by Clark (1998), in a framework that assumes two types of reactions, of fast and slow type (Jabari Kohpaei and Sathasivan, 2011).

## 3 Second order kinetic model

Starting from Clark (1998), the second order kinetic model is here written for a generic disinfectant and reacting compounds, producing generic *dbp*s. As previously reported, a complex set of parallel and serial, fast and slow, reactions determine the chemistry of a disinfectant decay (Kastl et al., 1999; Jabari Kohpaei and Sathasivan, 2011; Fisher et al., 2015; Wang et al., 2019; Monteiro et al., 2015). The use of a synthetic second order kinetic model is consistent with its use into a steady-state hydraulic analysis. In fact, steady-state hydraulic analysis means to capture the average system behaviour at a time scale from minutes to hours, varying the system boundary conditions (Giustolisi and Walski, 2012). Therefore, the chemistry of the disinfectant decay cannot be more detailed than hydraulic modelling. Consequently, the stoichiometries and reaction rate constants of reactants and *dbp*s should be intended representative of reactions complexity. The use of a second order kinetic model, being more chemically based than the first order one, provides the opportunity to estimate the stoichiometries and reaction rate constants for the specific system through laboratory experiments, as part of the water quality management framework.

In the following, *dis* and *rc* are used to indicate any kind of disinfectant and reacting compounds, respectively. The general chemical balanced reaction between a *dis* and *rc*, resulting in the formation of multiple *dbp*s ($P_i$), is the following:

$$a \cdot dis + b \cdot rc \leftrightarrow p_1 \cdot dbp_1 + \ldots + p_n \cdot dbp_n \qquad (2)$$

where $a$, $b$ and $p_1$, …, $p_n$ are the stoichiometric coefficients. As previously said, Eq. (2) summarizes the chemical complexity of disinfectant decay due to reacting compounds producing *dbp*s.

As reported by Clark (1998), the relationships between stoichiometric coefficients and reaction rate constants can be defined as follows:

$$k_{dis}/a = k_{rc}/b = k_{p1}/p_1 = \ldots = k_{pn}/p_n \qquad (3)$$

where $k_{dis}$, $k_{rc}$, and $k_{p1}$, …, $k_{pn}$ are the reaction rate constants (also named decay constants) for *dis*, *rc* and *dbp*s, respectively. It is useful to define $b_a$ and $p_{1a} \ldots p_{1n}$ as follows:

$$
\begin{aligned}
b_a &= \frac{b}{a} = \frac{k_{rc}}{k_{dis}} \\
p_{1a} &= \frac{p_1}{a} = \frac{k_{dbp1}}{k_{dis}} \\
&\dots \\
p_{1n} &= \frac{p_n}{a} = \frac{k_{dbpn}}{k_{dis}}
\end{aligned}
\tag{4}
$$

Allowing us to write:

$$
dis + b_a \cdot rc \leftrightarrow p_{1a} \cdot dbp_1 + \ldots + p_{na} \cdot dbp_n \tag{5}
$$

Therefore, Eq. (5) explicit the role of reaction rate constants in Eq. (2). Then, the second order kinetic model presented by Clark (1998) can be written by explicitly including the decay of reacting compounds and the formation of *dbp*s, as follows:

$$
\begin{aligned}
&\frac{dC_{dis}}{dt} = -k_{dis} \cdot C_{rc} \cdot C_{dis} = -k'_{dis} \cdot C_{dis} && k'_{dis} = f\left(C_{rc}\right) = k_{dis} \cdot C_{rc} \\
&\frac{dC_{rc}}{dt} = -k_{rc} \cdot C_{dis} \cdot C_{rc} = -k'_{rc} \cdot C_{dis} && k'_{rc} = f\left(C_{rc}\right) = k_{rc} \cdot C_{rc} \\
&\frac{dC_{p1}}{dt} = k_{p1} \cdot C_{dis} \cdot C_{rc} = k'_{p1} \cdot C_{dis} && k'_{p1} = f\left(C_{rc}\right) = k_{p1} \cdot C_{rc} \\
&\dots && \dots \\
&\frac{dC_{pn}}{dt} = k_{pn} \cdot C_{dis} \cdot C_{rc} = k'_{pn} \cdot C_{dis} && k'_{pn} = f\left(C_{rc}\right) = k_{pn} \cdot C_{rc}
\end{aligned}
\tag{6}
$$

Note that the reaction rate constants, *k*, of Eqs. (6) are model constants, but the same equations can be written in a first order kinetic model defining the reactions rates *k*' depending on the concentration of the reacting compounds. Therefore, using a chemically based second order kinetic model provides constant reaction rate constants, *k*, keeping in mind the synthetic modelling approach previously discussed. Note that *k*' = $f(C_{rc})$ in Eqs. (6) become variables in the first order kinetic model. The calibration value of such variables depends on $C_{rc}$ and the lack of its information does not allow the generalization of the results.

The differential equations in system (6), considering Eqs. (4), can be written as follows:

$$
\begin{aligned}
&\frac{dC_{dis}}{dt} = -k_{dis} \cdot C_{rc} \cdot C_{dis} \\
&\frac{dC_{rc}}{dC_{dis}} = \frac{k_{rc}}{k_{dis}} = b_a \\
&\frac{dC_{p1}}{dC_{dis}} = -\frac{k_{p1}}{k_{dis}} = -p_{1a} \\
&\ldots \\
&\frac{dC_{pn}}{dC_{dis}} = -\frac{k_{pn}}{k_{dis}} = -p_{na}
\end{aligned} \tag{7}
$$

Note that the first equation is the second order kinetic model for disinfectant decay, while the others are the ratios between the differential of the specific reacting compounds *rc* and by-products *dbp*s and that of disinfectant *dis*.

From Clark (1998), the solution of the first differential Eq. (7) is:

$$
C_{dis}\left(t, C_{dis}^0, C_{rc}^0, b_a, k_{dis}\right) = \frac{C_{dis}^0 (1-\alpha)}{1-\alpha \cdot e^{(\alpha-1)\cdot\beta\cdot t}} \quad \alpha = \frac{C_{rc}^0}{b_a \cdot C_{dis}^0}; \quad \beta = b_a \cdot k_{dis} \cdot C_{dis}^0 \tag{8}
$$

while the solutions of the other Eqs. (7) are:

$$
\begin{aligned}
&C_{rc}\left(t, C_{dis}^0, C_{rc}^0, b_a\right) - C_{rc}^0 = -b_a\left(C_{dis}^0 - C_{dis}(t)\right) \\
&C_{p1}\left(t, C_{dis}^0, C_{p1}^0, p_{1a}\right) - C_{p1}^0 = p_{1a}\left(C_{dis}^0 - C_{dis}(t)\right) \\
&\ldots \\
&C_{pn}\left(t, C_{dis}^0, C_{pn}^0, p_{na}\right) - C_{pn}^0 = p_{na}\left(C_{dis}^0 - C_{dis}(t)\right)
\end{aligned} \tag{9}
$$

Therefore, the Eq. (8) states that the concentration decay of a disinfectant depends on the initial concentration of the disinfectant itself and of the reacting compounds, $C^0_{dis}$ and $C^0_{rc}$, respectively, the stoichiometric ratio between reacting compounds and disinfectant, $b_{\mathrm{a}}$, and the reaction rate constant of the disinfectant, $k_{dis}$. Note that $b_{\mathrm{a}}$ and $k_{dis}$ are characteristic, although synthetic, of the specific chemical reaction. The Eqs. (9) state that the reacting compounds decay and the *dbp*s growth depend linearly on the reduction of disinfectant concentration by $b_{\mathrm{a}}$ and $p_{ia}$, respectively.

The effectiveness of using the second order kinetic model can be demonstrated referring to a technical situation due to the standard policy for reservoir management of controlling the organic compounds. In that case, $C^0_{rc} = 0$ and $\alpha = 0$ and, therefore, the solutions to Eqs. (8) and (9) are:

$$
\begin{aligned}
&C_{dis}\left(t, C_{dis}^0, C_{rc}^0 = 0, b_a, k_{dis}\right) = C_{dis}^0 \\
&C_{p1} = C_{p1}^0, \quad \ldots, C_{pn} = C_{pn}^0
\end{aligned} \tag{10}
$$

Note that the model correctly represents a normal situation of water quality policy at reservoirs with no disinfectant decay and no *dbp*s growth, indicating that, if a disinfectant decay is observed, this should be due to causes internal to the hydraulic system. This introduces the need to define the endogenous mechanism of disinfectant decay in the proposed framework, since aged pipes are characterized by wall biofilm that could also be detached by the momentum of the water flow. Note that the first order kinetic model cannot explain the endogenous reason of disinfectant decay because information about reacting compounds is absent, as clarified in the following paragraph.

## 4 Endogenous mechanism of disinfectant decay

To the purpose of the following discussion, Eq. (8) and Eqs. (9) are rewritten as follows:

$$\begin{aligned}
&C_{dis}\left(t, C_{dis}\left(t-\Delta t_q\right), C_{rc}\left(t-\Delta t_q\right), b_a, k_{dis}\right) = \frac{C_{dis}\left(t-\Delta t_q\right)\left(1-\alpha\left(t-\Delta t_q\right)\right)}{1-\alpha\left(t-\Delta t_q\right)\cdot e^{\left(\alpha\left(t-\Delta t_q\right)-1\right)\cdot\beta\left(t-\Delta t_q\right)\cdot t}} \\
&\alpha\left(t-\Delta t_q\right) = \frac{C_{rc}\left(t-\Delta t_q\right)}{b_a \cdot C_{dis}\left(t-\Delta t_q\right)}; \quad \beta\left(t-\Delta t_q\right) = b_a \cdot k_{dis} \cdot C_{dis}\left(t-\Delta t_q\right)
\end{aligned} \tag{11}$$

and

$$\begin{aligned}
&C_{cr}\left(t, C_{dis}\left(t-\Delta t_q\right), C_{rc}\left(t-\Delta t_q\right), b_a\right) = C_{rc}\left(t-\Delta t_q\right) - b_a\left(C_{dis}\left(t-\Delta t_q\right) - C_{dis}(t)\right) \\
&C_{p1}\left(t, C_{dis}\left(t-\Delta t_q\right), C_{p1}\left(t-\Delta t_q\right), p_{1a}\right) = C_{p1}\left(t-\Delta t_q\right) + p_{1a}\left(C_{dis}\left(t-\Delta t_q\right) - C_{dis}(t)\right) \\
&\dots \\
&C_{pn}\left(t, C_{dis}\left(t-\Delta t_q\right), C_{pn}\left(t-\Delta t_q\right), p_{na}\right) = C_{pn}\left(t-\Delta t_q\right) + p_{na}\left(C_{dis}\left(t-\Delta t_q\right) - C_{dis}(t)\right)
\end{aligned} \tag{12}$$

where $\Delta t_q$ is the time step of the water quality analysis. In fact, the Lagrangian diffusion scheme into networks is based on parcels in each pipe moving in the flow direction at each time step of the analysis $\Delta t_q$ (Giustolisi et al., 2023). In such Lagrangian scheme, the reaction occurs every $\Delta t_q$, and the initial concentrations $C^0$ for the calculation at time $t$ of the reactions are those at $t - \Delta t_q$.

As said in the introduction and literature review paragraphs, the wall of aged pipes in WDNs is characterized by a biofilm layer having a complex structure composed of various microbial species (Bendinger et al., 1993; Hallam et al., 2002; Bonadonna et al., 2008). Wall biofilm develops especially where the flow rate is low or stagnation conditions occur (e.g., branches, terminal sections towards users, taps, etc.). Moreover, high pipe roughness and irregular textures provide more niches for microbial attachment and protection (LeChevallier et al., 1988). Additionally, microorganisms attach more rapidly to hydrophobic and non-polar surfaces, such as Teflon and other plastics, rather than to hydrophilic materials such as vitrified clay or metals. Some studies have demonstrated that plastic materials, such as cross-linked polyethylene (PE-Xb) (Bonadonna et al., 2008), represent more suitable sites for biofilm formation than metallic materials like steel and copper (Lehtola et al., 2004). Once formed, the wall biofilm supports microbial survival and, under favourable conditions, its growths (Bonadonna et al., 2008).

The disinfectant decay originates by the contact at the pipe wall biofilm and by that biofilm compounds that could be detached due to the momentum of water flow (Lehtola et al., 2006; Tsai, 2005). For this reason, it is here presented a novel modelling framework for these two different endogenous mechanisms causing two distinct disinfectant decay patterns into the networks, referable as wall and bulk decays.

First, the parameters $\alpha$ and $\beta$ of Eq. (11) needs to be rewritten as follows:

$$\alpha = \frac{C_{rc}\left(t-\Delta t_q\right) + C_{rc-bulk}\left(t-\Delta t_q\right) + C_{rc-wall}\left(t-\Delta t_q\right)}{b_a \cdot C_{dis}\left(t-\Delta t_q\right)}$$
$$\beta = b_a \cdot \left(k_{dis-bulk} + k_{dis-wall}\right) \cdot C_{dis}\left(t-\Delta t_q\right) \qquad (13)$$

where $C_{rc}$, $C_{rc-wall}$, $C_{rc-bulk}$ are, at the previous step $t - \Delta t_q$ (initial condition for $t$), the concentration of the reacting compounds in the bulk of the water coming from the upstream parcel and the endogenous reacting compounds at wall biofilm and detached going in the bulk, respectively. Consequently, it is necessary to define two reaction rate constants, $k_{dis-wall}$ and $k_{dis-bulk}$, because the velocity of reactions at the wall and bulk can be different. In fact, it is reasonable to assume that $k_{dis-bulk}$ is, even one order of magnitude lower than $k_{dis-wall}$ (Jabari Kohpaei and Sathasivan, 2011).

In fact, the reacting compounds are characterized by different reaction rate constants: at the pipe wall, they are assumed to be higher because the contact between disinfectant and reacting compounds is local and not persistent (i.e., the compounds that react with the disinfectant passing with the water flow are always the most ready to react and therefore are those with higher reaction rate constants). The biofilm compounds that detach upon entering the bulk water, despite being both fast and slow reacting, have more time to react with the disinfectant because their contact is persistent, as the disinfectant flows downstream with the water velocity, and they are, therefore, assumed to have lower reaction rate constants (Wang et al., 2019).

Now the issue is to define $C_{rc-wall}$ and $C_{rc-bulk}$; to the purpose it is necessary to define the masses of $C_{rc-wall}$ and $C_{rc-bulk}$, $m_{rc-wall}$ and $m_{rc-bulk}$, respectively, that are expressed in grams per day per square meter of internal pipe wall [g/m$^2$/d]. Consequently, the following equations can be written at pipe level:

$$C_{rc-wall}\left(T_p\right)\left[\frac{g}{m^3}\right]=m_{rc-wall}\left[\frac{g}{m^2\cdot d}\right]\cdot\frac{S_p\left[m^2\right]}{V_p\left[m^3\right]}\cdot T_p\left[d\right]=m_{rc-wall}\cdot\frac{4}{D_p}\cdot T_p$$

$$C_{rc-bulk}\left(L_p\right)\left[\frac{g}{m^3}\right]=\left\{m_{rc-bulk}\left[\frac{g}{m^2\cdot d}\right]\cdot\frac{S_p\left[m^2\right]}{V_p\left[m^3\right]}\cdot T_p\left[d\right]\right\}\cdot v_p=m_{rc-bulk}\cdot\frac{4}{D_p}\cdot L_p \tag{14}$$

where $D_p$ [m] is the pipe internal diameter, $S_p$ [$m^2$] and $V_p$ [$m^3$] are respectively the internal pipe surface and the internal pipe volume, $L_p$ [m] is the pipe length and $v_p$ [m/s] is the average velocity of the hydraulic analysis. Therefore, considering the Lagrangian scheme here adopted, the Eqs. (14) can be written at parcel level as:

$$C_{rc-wall}\left(\Delta t_q\right)=m_{rc-wall}\cdot\frac{4}{D_p}\cdot\Delta t_q$$

$$C_{rc-bulk}\left(dx_p\right)=m_{rc-bulk}\cdot\frac{4}{D_p}\cdot dx_p \tag{15}$$

where $dx_p$ [m] is the length of the parcel and $\Delta t_q$ [d] is the water quality step. As shown in Eqs. (14) or (15), the concentration of reacting compounds at wall, $C_{rc\text{-}wall}$, given a certain daily mass $m_{rc\text{-}wall}$, increases for small diameters and for slow flows, i.e., taking high $T_p$ to travel the pipe (e.g., distribution pipes that feed the consumers or branched pipes). Similarly, it is worthy to observe that the concentration of reactive compounds that detach from the wall and move into the pipe bulk water, $C_{rc\text{-}bulk}$, for a certain daily mass $m_{rc\text{-}bulk}$, increases for small diameters and for very long pipes, being, moreover, the result of the progressive sum of the contributions of each parcel along the development of the WDN. Note that $k_{dis\text{-}wall} = 0$ or $k_{dis\text{-}bulk} = 0$ in Eq. (13) if $C_{rc\text{-}wall}(t - \Delta t_q) = 0$ or $C_{rc\text{-}bulk}(t - \Delta t_q) = 0$, respectively. Furthermore, *dbp*s need to be computed considering the disinfectant consumption due to $C_{rc\text{-}wall}(t - \Delta t_q)$ and/or $C_{rc\text{-}bulk}(t - \Delta t_q)$, while the decrease of the reacting compounds refers to bulk reactions only, i.e. $C_{rc} + C_{rc\text{-}bulk}$ of transported and detached ones, respectively.

Finally, private and public water storage tanks are also crucial elements for water quality. Some studies (van der Merwe et al., 2013; Javed et al., 2025) have investigated the biofilm formation in water tanks concluding that it is affected by bacterial growth and proliferation. The need to maintain adequate residual chlorine levels to inhibit biofilm formation was emphasized. The use of materials with smoother and more regular surfaces, possibly with antimicrobial coatings, helps to reduce biofilm adhesion on both plastic and concrete tanks, although the regular cleaning and maintenance of tanks is mandatory.

Eq. (15) for tanks can be written as follows:

$$C_{rc-tank}\left(\Delta t_q\right) = m_{rc-tank} \cdot \left(\frac{4}{D_{tank}} + \frac{1}{H}\right) \cdot \Delta t_q \tag{16}$$

where $C_{rc\text{-}tank}$ is the concentration of reacting compounds in $\Delta t_q$, $D_{tank}$ is the diameter of the tank (assuming a cylindric shape), $4/D_{tank}$ is its hydraulic radius and $H$ is the water level in the tank. Note that $m_{rc\text{-}tank}$ refers to the contribution of the reacting compounds at wall and detached. It is worth noting that splitting $m_{rc\text{-}tank}$ in wall and detached reacting compounds is useful to assess their increase in the bulk influencing the water quality into the network or at the consumers (private tanks).

# 5 Analysis tools for water distribution networks

## *5.1 Advanced steady-state hydraulic analysis*

As reported in Giustolisi et al. (2008; 2024) the mathematical model for steady-state hydraulic analysis of a pressurized water pipeline system composed of $n_p$ pipes with unknown flow rates, $n_n$ nodes with unknown heads and $n_0$ nodes with known heads is expressed by two set of equations of momentum and mass balance at pipes and nodes, respectively, as follows:

$$\begin{array}{lcl} \mathbf{A}_{pp}\mathbf{Q}_p\left(t,\Delta t_h\right) + \mathbf{A}_{pn}\mathbf{H}_n\left(t,\Delta t_h\right) & = & -\mathbf{A}_{p0}\mathbf{H}_0\left(t,\Delta t_h\right) \\ \mathbf{A}_{np}\mathbf{Q}_p\left(t,\Delta t_h\right) - \mathbf{d}_n\left(t,\mathbf{H}_n\left(t,\Delta t_h\right)\right) & = & \mathbf{0}_n \\ \mathbf{d}_n\left(\mathbf{H}_n\left(t,\Delta t_h\right)\right) = \dfrac{\mathbf{V}_n\left(\mathbf{H}_n\left(t,\Delta t_h\right)\right)}{\Delta t_h} & & \end{array} \tag{17}$$

where $\mathbf{A}_{pp}$=[$n_p$, $n_p$] is a diagonal matrix with elements based on the pipes resistance; $\mathbf{Q}_p$=[$n_p$, 1] is a column vector of unknown pipes flow rate; $\mathbf{H}_n$=[$n_n$,1] is a column vector of unknown nodal heads; $\mathbf{H}_0$=[$n_0$, 1] is a column vector of known nodal heads; $\mathbf{0}_n$=[$n_n$, 1] is a column vector of null values; $\mathbf{d}_n$=[$n_n$, 1] is a column vector of nodal water flows; $\mathbf{A}_{pn}$= $\mathbf{A}_{np}^T$ and $\mathbf{A}_{p0}$ are the topological incidence sub-matrices of size [$n_p$, $n_n$] and [$n_p$, $n_0$], respectively, derived from the general topological matrix $\bar{\mathbf{A}}_{pn}$=[$\mathbf{A}_{pn} \mid \mathbf{A}_{p0}$] of size [$n_p$, $n_n$+ $n_0$] and $\mathbf{d}_n$ is the mean value of each nodal demand component during the time interval $t+\Delta t_h$, $\Delta t_h$ is the hydraulic timestep representative of the model and $\mathbf{V}_n$=[$n_n$,1] is a column vector of the related volumes. Consistently, the status variables $\mathbf{Q}_p$ and $\mathbf{H}_n$ do not represent instantaneous values but, in good technical approximation, the mean values over $t + \Delta t_h$.

The hydraulic model in system (17) is the general form for pressure-driven analysis (PDA) (Giustolisi et al., 2008) meaning that the hydraulic status of the water system, $\mathbf{Q}_p$ and $\mathbf{H}_n$, is driven by pressure dependency of nodal demands, $\mathbf{d}_n$. This is a more general assumption of the demand-driven analysis (DDA) assuming the priors about $\mathbf{d}_n$. Note that $\mathbf{d}_n$ represents the summation of different components of demands as for example customer and leakage ones (Giustolisi and Walski, 2012).

The system in (17) is non-linear with respect to both momentum and mass balance equations whose status variables, $\mathbf{Q}_p$ and $\mathbf{H}_n$, and boundary conditions are the representative values for the assumed steady-state timestep, $\Delta t_h$, depending on the analysis purpose.

Giustolisi et al. (2024; 2008) introduced the leakage model component, using the hydraulic analysis at pipe level, that can be performed using (i) the Power model (Germanopoulos, 1985) or (ii) the Fixed Area and Variable Area Discharge (FAVAD) model (May, 1994; van Zyl and Cassa, 2014).

### *5.2 The Lagrangian diffusion scheme*

The water quality analysis is here performed using the velocity field calculated using Eqs. (17). The Lagrangian method is here used to model the diffusion into the network of the reacting species. It requires the assumption of a steady-state quality timestep, $\Delta t_q$, that must be lower than $\Delta t_h$. The Lagrangian scheme is based on parcels moving downstream of the pipes at each water quality step $\Delta t_q$. The number of parcels depends on the velocity at $\Delta t_h$, the pipe length and $\Delta t_q$; therefore, the number of parcels varies with pipes and increases with lower $\Delta t_q$ (Giustolisi et al., 2023). The accuracy of the Lagrangian scheme is, then, controlled by the maximum allowed number of parcels for long length and/or low velocity pipes and by $\Delta t_q$ for short length and/or high velocity pipes.

To the purpose, a strategy based on programming with cell matrices data type allows overcoming the issue of the maximum number of parcels (Giustolisi et al., 2023) considerably increasing the water quality analysis accuracy. The use of cell data matrix does not constrain the size of a single matrix that is used to store data of all the parcel in a classic programming approach. In fact, the maximum amount of data stored for parcels depends on the summation of the maximum number of parcels over all pipes, thus the constraint becomes the CPU memory and the indexing capacity, more details can be found in Giustolisi et al. (2023).

## 6 Small-sized network and patterns of the two endogenous decay mechanisms

The small-sized network of Monteparano, managed by Acquedotto Pugliese, supplies about 2,000 inhabitants. The hydraulic model of the system is reported in Figure 1. The network is composed of 182 pipes and 160 nodes, with total length of about 16 km and pipe diameters ranging from 200 mm (transport lines from the reservoir) to 60 mm (peripheral lines). A single reservoir feeds 977 consumer meters.

The hydraulic model was calibrated using real data provided by Acquedotto Pugliese. The mean velocities are low in the looped part of the network and peripherical lines, while they are about 1 m/s in the transport lines.

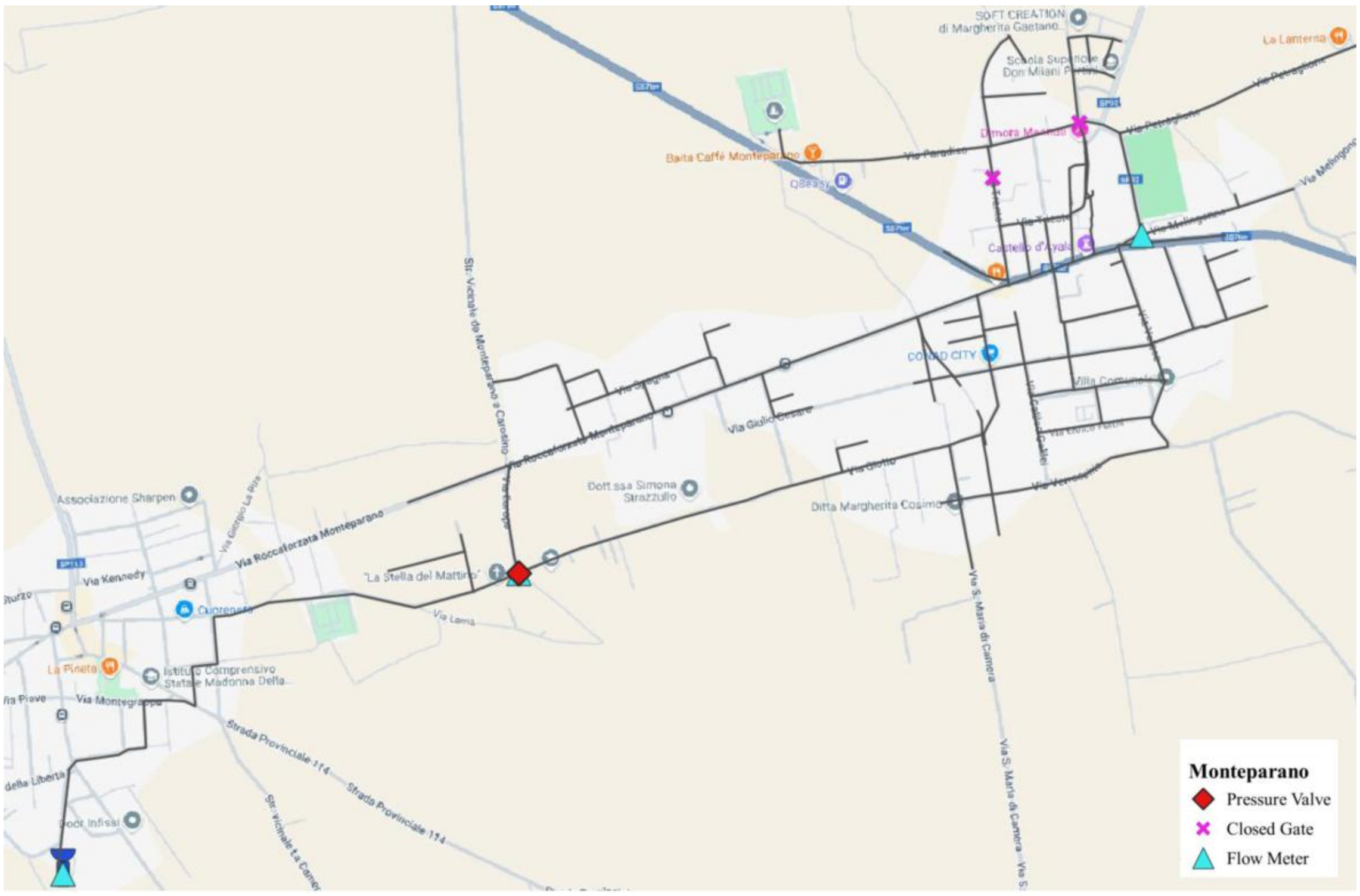

Figure 1. Water distribution network of Monteparano.

As previously reported, the standard policy of reservoir management is to control the organic compounds generation, i.e., $C_{rc} = 0$ is the boundary condition at the reservoir. Despite the initial condition at the reservoir, due to the water quality policy, dosing a disinfectant at the reservoir is a guarantee for water quality. In fact, it is well-known by technicians that boundary conditions at pipe wall can cause disinfections decay, i.e., the endogenous decay mechanisms. Without impairing the generality of the discussion, the continuous dosing of disinfectant at the reservoir is here assumed equal to 0.5 g/m$^3$.

To demonstrate the two endogenous decay mechanisms generating two distinct patterns, two scenarios have been tested:

- Wall scenario: detachment from the biofilm due to the water velocity does not occur and the disinfectant reacts with biofilm, only. It is assumed a disinfectant reaction rate constant $k_{dis\text{-}wall}$ = 10 [m$^3$/g/d], thus assuming that at the pipe wall fast reactions mainly happen (Jabari Kohpaei and Sathasivan, 2011). The mass of reacting compounds at the wall was assumed $m_{rc\text{-}wall}$ = 10 [g/m$^2$/d].
- Bulk scenario: detachment from the biofilm due to the water velocity generates the reacting compounds into the bulk water. Note that it is an instrumental assumption to separate the contributions of the two mechanisms to the purpose of the case study. It is assumed a disinfectant reaction rate constant $k_{dis\text{-}bulk}$ = 0.5 [m$^3$/g/d], thus in the bulk water slow reactions mainly happen. The mass of reacting compounds into the bulk was assumed $m_{rc\text{-}bulk}$ = 1 [g/m$^2$/d] per unit velocity (Jabari Kohpaei and Sathasivan, 2011).

Without impairing the generality of the results, the stoichiometric ratio of the reacting compounds with respect to chlorine, $b_a$, was set equal to 0.25 (Kastl et al., 1999).

Figure 2 shows the average water age of a standard operating cycle of Monteparano WDN. It can be observed that most of the nodes are characterised by a water age far below 24 hours (approximately 8 hours) and nineteen nodes, the peripheral ones, are characterised by a water age above 48 hours.

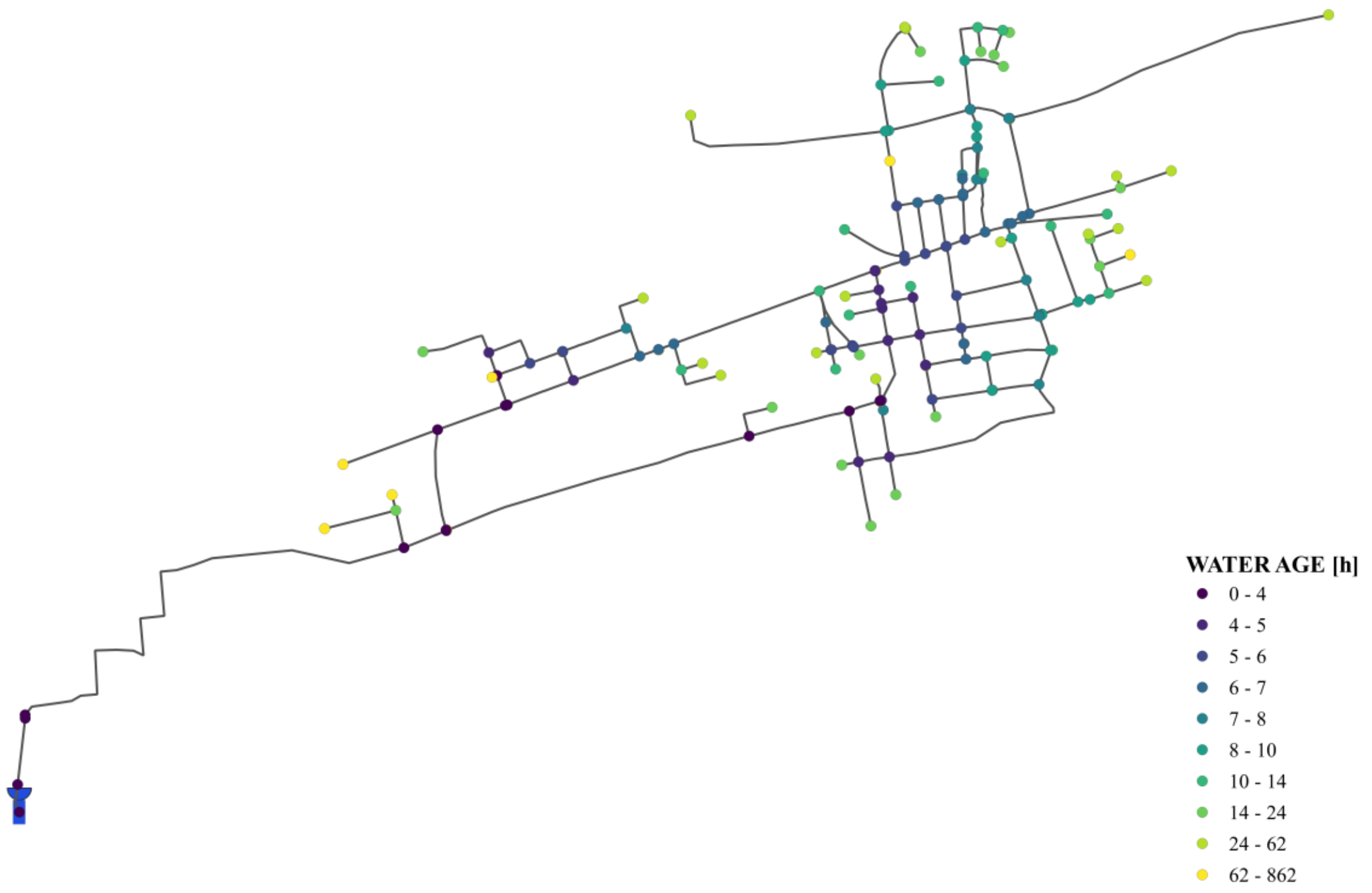

Figure 2. Water age analysis of the Monteparano network

Figure 3 reports the pattern of the endogenous mechanism due to disinfectant reaction with wall biofilm. From Figure 3, mean nodal disinfectant concentrations is not relevantly low at the peripheral nodes with respect to the internal ones, therefore, the disinfectant decay is not much correlated with the average water age. Note that $T_p$ in Eq. (14) refers to the local contribution of the single pipe, but the shortest path at network level determining the water age generally do not pass through the pipes with higher $T_p$. In any case, Eq. (14) indicates that oversized transport lines are threatening for water quality due to residence time causing a relevant wall decay.

Figure 4 reports the pattern of the endogenous mechanism due to disinfectant reaction with detached wall biofilm. From Figure 4, mean nodal disinfectant concentrations is relevantly lower at the peripheral nodes with respect to the internal ones, therefore, the disinfectant decay is much correlated with the average water age. Note that at the network level the release of reacting compounds detached from the wall and going into the bulk water contribute progressively along the shortest paths to the disinfectant decay so that it is relevant at the peripheral nodes and, thus, correlated with water age.

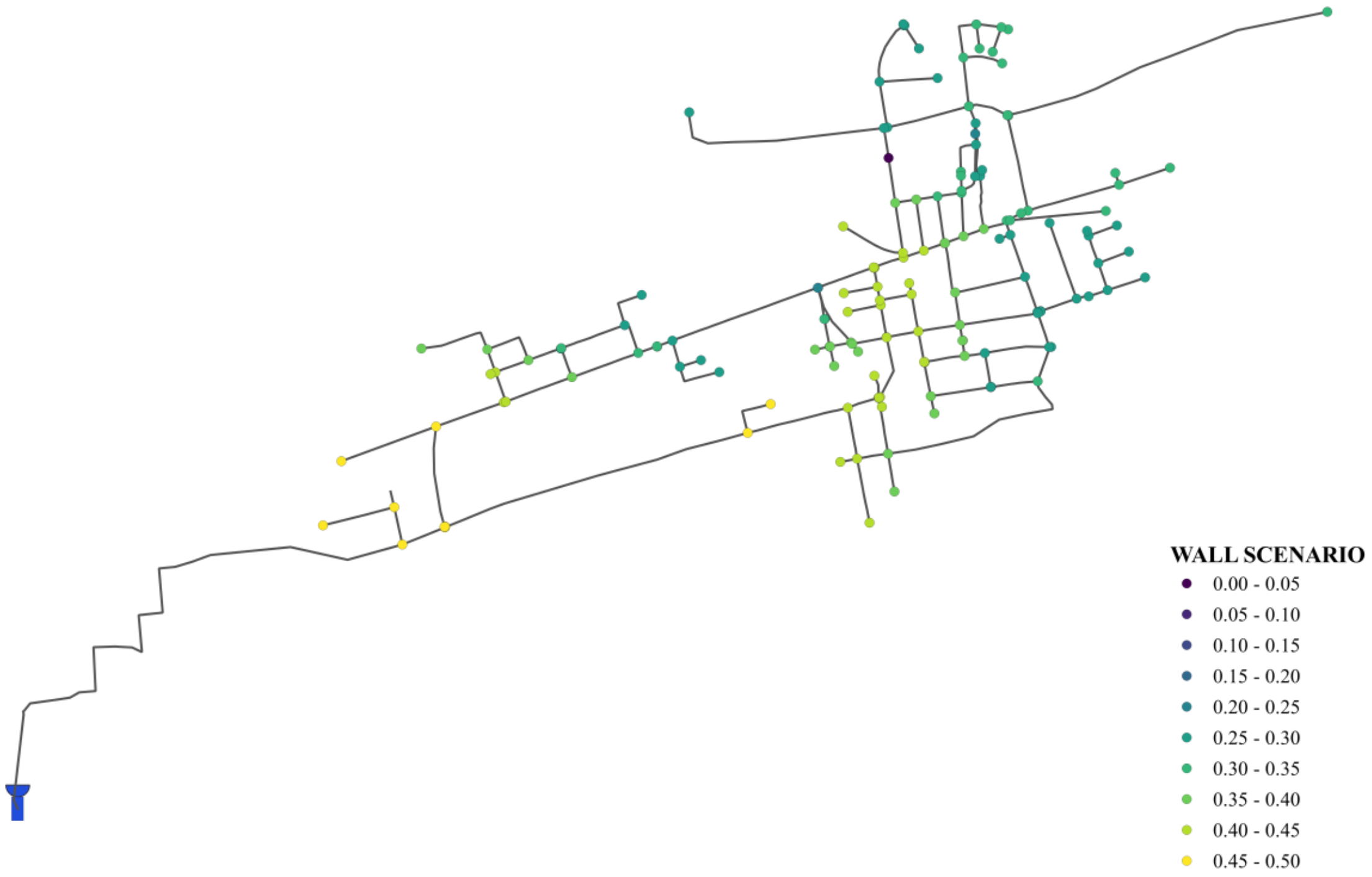

Figure 3. Mean nodal disinfectant concentrations in [g/m$^3$]: Wall scenario.

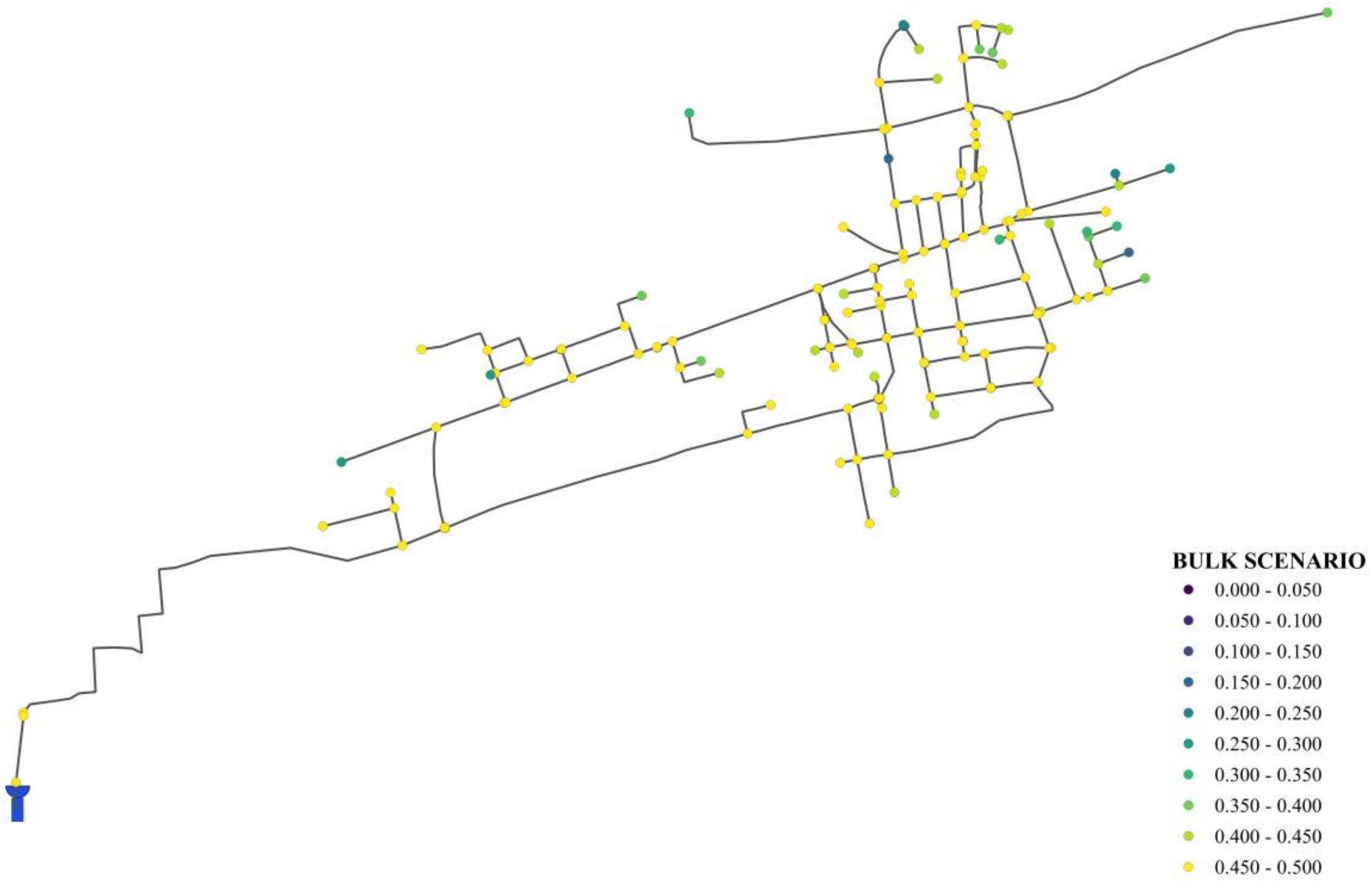

Figure 4. Mean nodal disinfectant concentrations in [g/m$^3$]: Bulk scenario.

The proposed framework proved to be effective for modelling of endogenous compounds due to biofilm at wall. It allows to distinguish between disinfectant consumption due to the two different mechanisms (wall biofilm detachment or not).

The novel framework has the potential to support the water utility for quality management based on risk assessment, for decisions relating to actions both with a short-term time horizon (emergencies, boosters) (Ohar and Ostfeld, 2014), and with longer-term time horizons (planning network cleaning, use of appropriate materials, pipe rehabilitation, etc.).

## 7 Large-sized network and disinfectant management using the novel framework

The aim of case study is to show and discuss the effectiveness of the proposed framework for water quality management considering the asset condition change due to management interventions.

To the purpose, the system of Bari, managed by Acquedotto Pugliese, is here used.

It supplies about 200,000 inhabitants, by means of three reservoirs feeding 16,981 consumer meters. The WDN is reported in Figure 5. The hydraulic model is composed of 5,286 pipes and 4,523 nodes, and the total length is about 407 km. Pipe diameters ranges from 1000 mm (transport lines from the reservoirs) to 40 mm (peripheral lines).

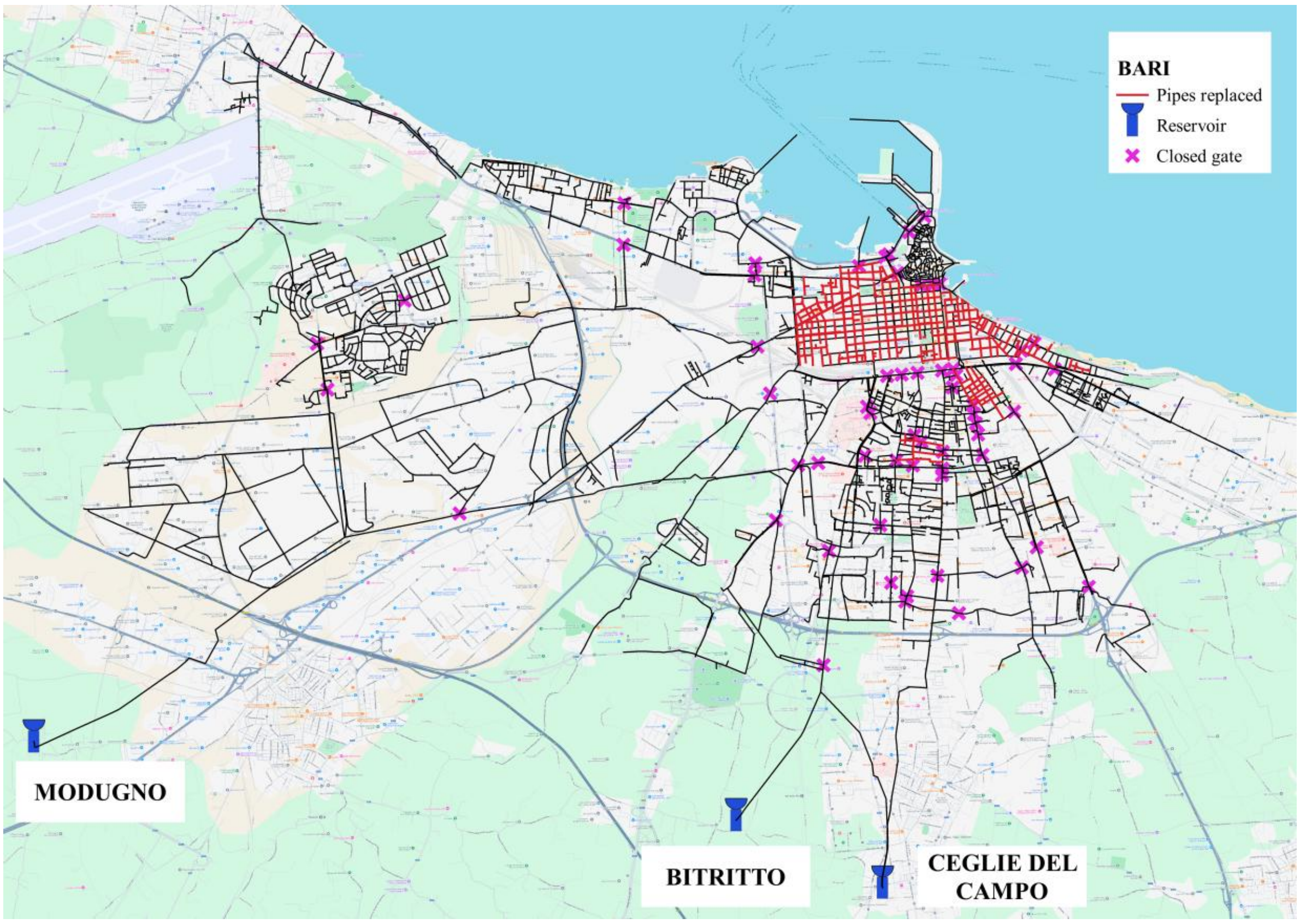

Figure 5. Water distribution network of Bari.

Figure 5 reports eight pressure control valves and the closed gates at the boundaries of thirty-four DMAs functional to water loss management. As part of the water loss management there is pipe replacement in the centre of the town, see label “pipe replaced” in Figure n.5.

The standard policy of reservoir management is to control the organic compounds generation, i.e., $C_{rc} = 0$ is the boundary condition at any reservoir as previously reported for Monteparano WDN.

The WDN of Bari is not new, therefore, the assumption of aged pipes is technically consistent, thus the two endogenous mechanisms of disinfectant decay, previously discussed, occur.

Consistently with the previously reported discussion and the technical-scientific literature (Jabari Kohpaei and Sathasivan, 2011) the bulk and wall reaction rate constants of the disinfectant (chlorine as in Monteparano case study) have been assumed $k_{dis\text{-}bulk} = 0.5$ [m$^3$/g/d] and $k_{dis\text{-}wall} = 10$ [m$^3$/g/d], respectively. The stoichiometric ratio of the reacting compounds with respect to chlorine, $b_a$, was set equal to 0.25 (Kastl et al., 1999). Note that the novel framework, as previously reported, aims at demonstrating the effectiveness of using the second order kinetic synthetic model and, consequently, to characterize the specific system through $b_a$ and reaction rate constants. For this reason, the stoichiometric ratio $b_a$ is not still available. The common use of first order kinetic model in WDNs make also not available studies on $b_a$ at system scale.

For the same reason about the novelty of the proposed framework, in Eqs. (14) or Eqs. (15), the values $m_{rc\text{-}bulk} = 0.5$ [g/m$^2$/d] for unit velocity and $m_{rc\text{-}wall} = 5$ [g/m$^2$/d] were set.

Without impairing the generality of the discussion, the continuous dosing of disinfectant at the reservoirs is here assumed equal to 0.7 g/m$^3$.

To show the effectiveness of the proposed framework two asset management activities, pipe replacement and cleaning of transmission lines, have been assessed with respect to water quality management. Therefore, the following water quality analyses, using the previously reported parameters, have been performed.

A. No asset management.
B. Pipe replacement. The water quality analysis has accounted for replacement of 1,000 pipes (about 42 km of total length, see label “pipe replaced” in Figure n.5). For those pipes $m_{rc\text{-}wall}$ and $m_{rc\text{-}bulk}$ are set null, meaning that there is not a biofilm contributing to disinfectant decay due to the two previously reported endogenous mechanisms.
C. Management of transmission lines. The water quality analysis has accounted for management of pipes with diameters greater than 300 mm, which are the transmission lines for the specific WDN system, by means of a cleaning activity aimed at removing the biofilm. For those pipes, assuming a complete removal of the biofilm, $m_{rc\text{-}wall}$ and $m_{rc\text{-}bulk}$ are set null. As in the previous case, there is no biofilm contribution to disinfectant decay due.

D. Both the previous management activities have been performed.

The Figures 6-9 show the results of the previous analyses in terms of nodal mean concentration of the disinfectant (chlorine) over a standard operative cycle.

Figure 6 shows a relevant disinfectant decay (low disinfectant concentration) in the ending nodes of terminal pipes into the network and in the coastal zone far from the reservoirs. The low disinfectant concentration is related to the two endogenous mechanisms of decay being $C_{rc} = 0$ at the reservoirs for water quality policy.

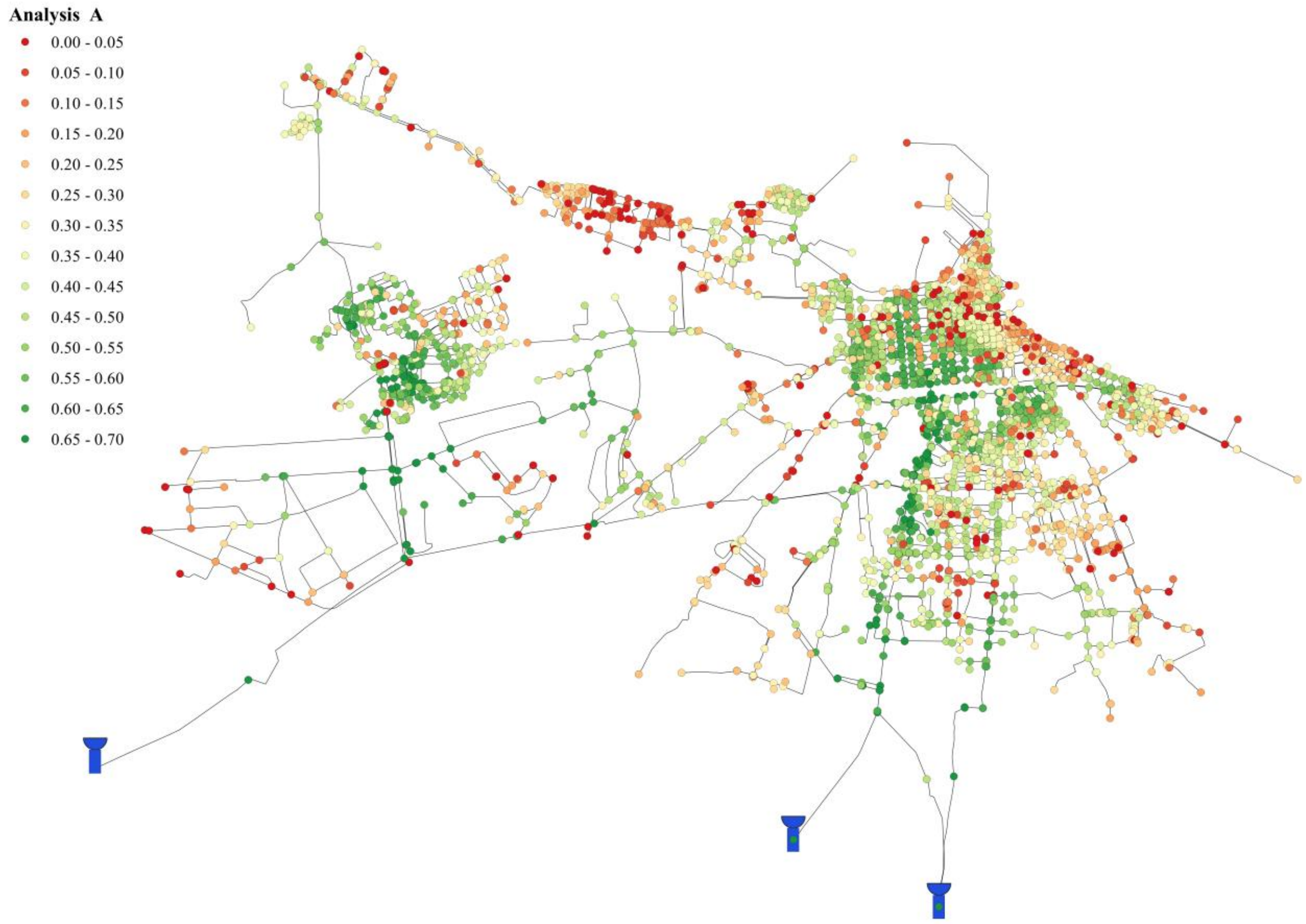

Figure 6. Mean nodal disinfectant concentrations in [g/m3]: Water quality analysis A.

Figure 7, referring to pipe replacement, shows that the absence of biofilm locally reduces the disinfectant decay compared to Figure 6. The effect is local because the nodes in the coastal area are generally not downstream of the area of replacements. The coastal area is still affected by the water quality of transmission lines. In fact, the bulk detachment in those long pipes (see the second of Eqs. 14) continues influencing the progressive and not local decay.

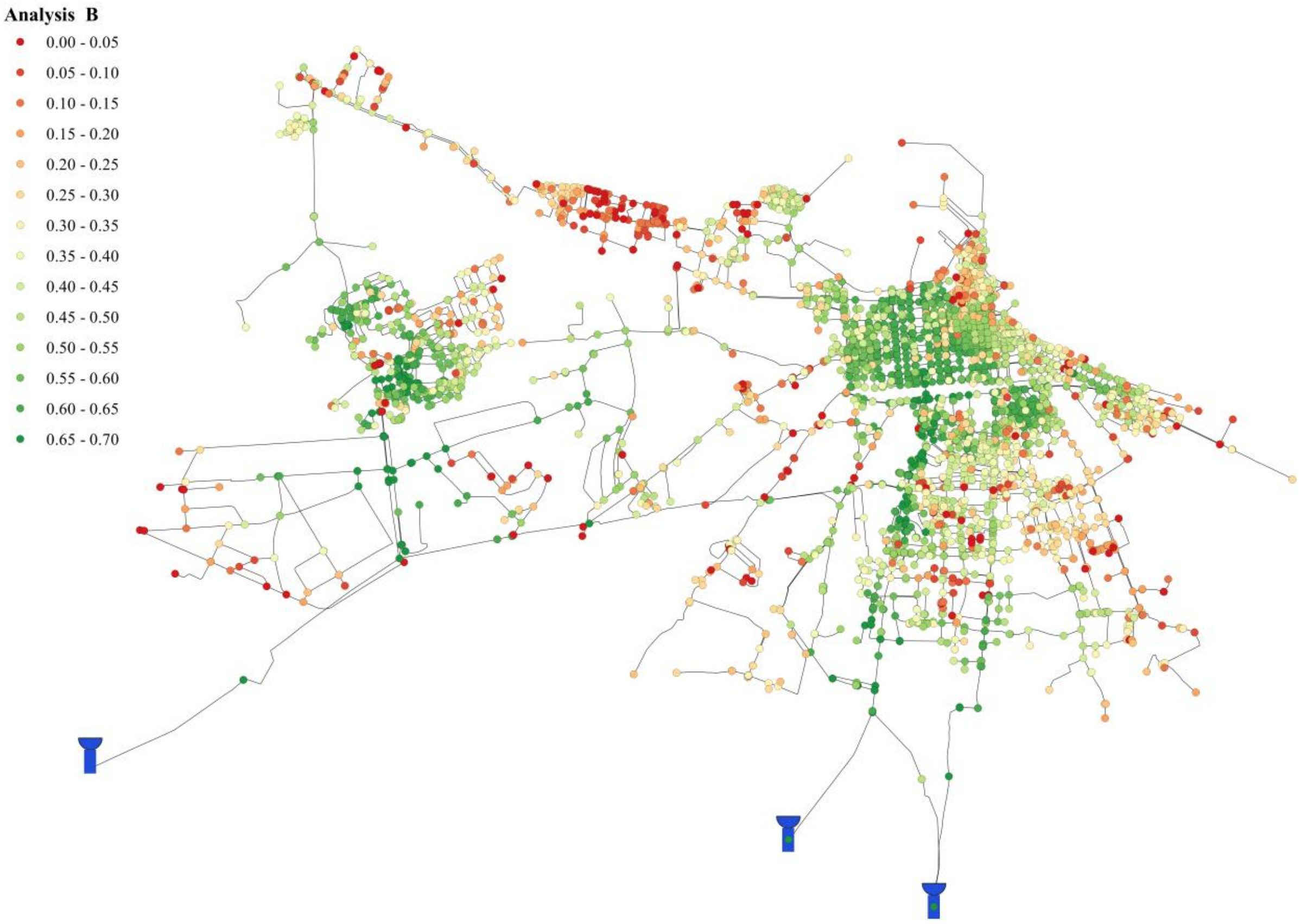

Figure 7. Mean nodal disinfectant concentrations in [g/m$^3$]: Water quality analysis B.

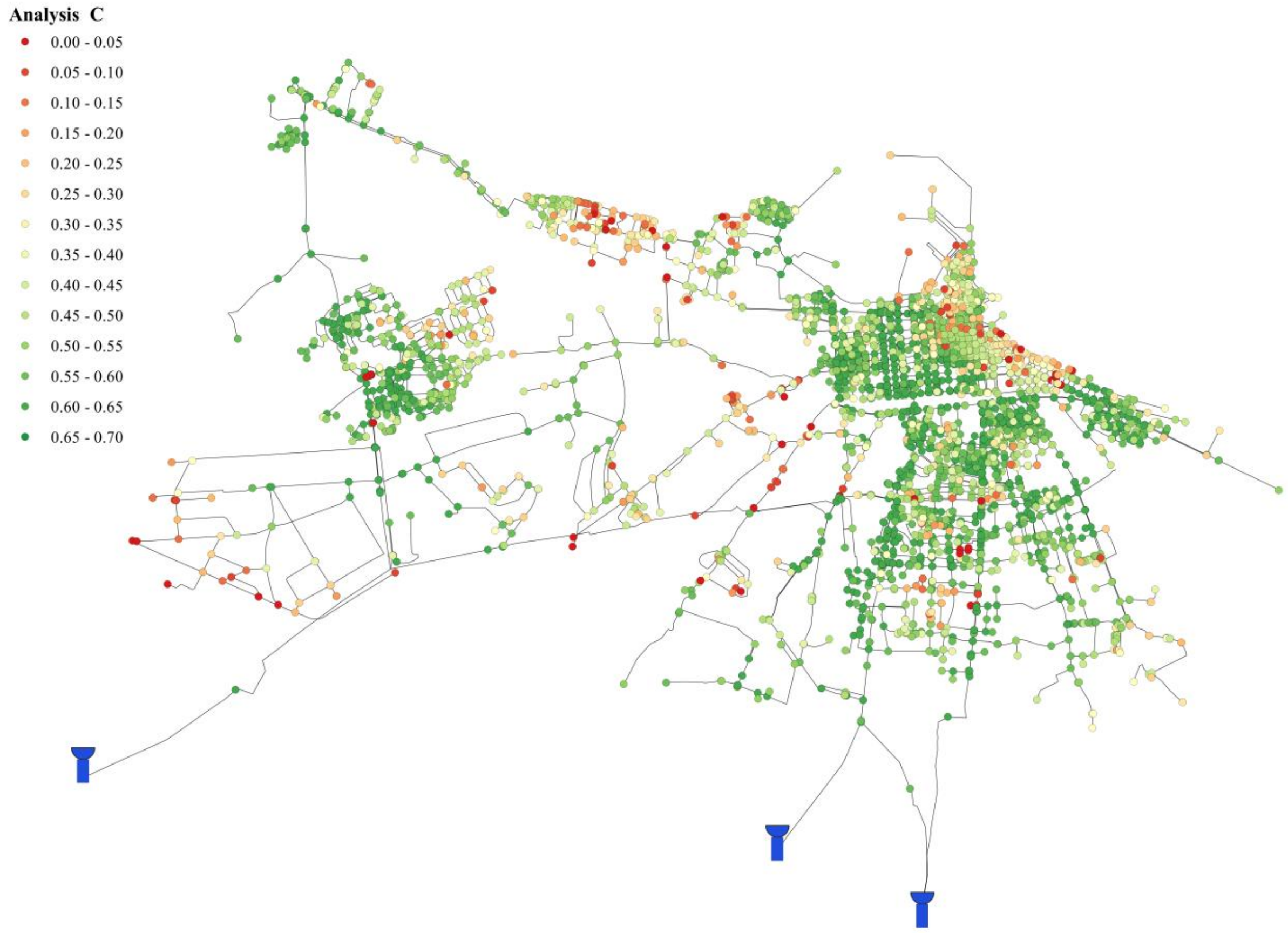

Figure 8. Mean nodal disinfectant concentrations in [g/m$^3$]: Water quality analysis C.

Figure 8, referring to management, e.g., cleaning, of transmission lines, shows that the absence of biofilm globally reduces the disinfectant decay compared to Figure 6. The effect is global because the nodes of the coastal area are downstream of the reservoirs and the progressive detachment along the transmission lines of the biofilm from the wall is absent together with the progressive disinfectant (chlorine) decay. Water quality issue remains at the ending node of terminal pipes, and the framework suggests local cleaning, being the decay caused by very low velocities (see first of Eqs. (14)). Overall, the present analysis highlights that the water quality management of the transmission lines is the most important but not a resolution for local situations.

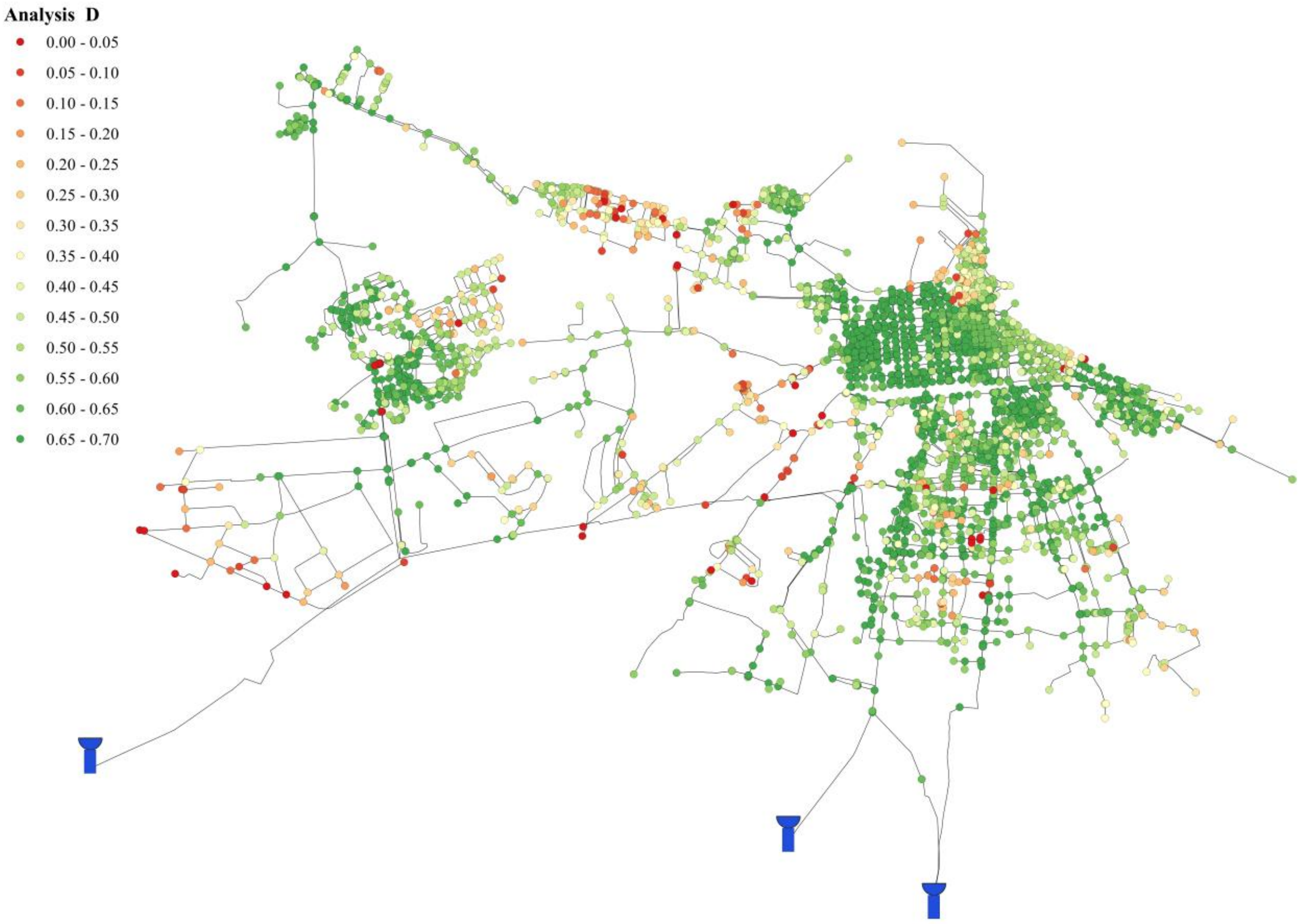

Figure 9. Mean nodal disinfectant concentrations in [g/m$^3$]: Water quality analysis D.

Finally, Figure 9 shows the analysis using the two previous (B and C) asset management activities. Those activities resolve the issue of disinfectant decay but not that one of the local situations related to the ending node of terminal pipes.

The previous analyses do not depend on the specific parameters of the endogenous mechanisms or second order kinetic synthetic model. The analyses demonstrate that the novel proposed framework, together with a monitoring system of the relevant parameters of water quality, e.g., the chlorine concentration, could be effective to support water quality management activities to achieve the quality of the public expenditure. Furthermore, the analyses framework can support the design of the monitoring system and the calibration of $m_{rc\text{-}wall}$ and $m_{rc\text{-}bulk}$.

## 8 Concluding remarks and future perspectives

The proposed framework for water quality modelling in WDNs integrates chemically based second order kinetic models for disinfectant and reactive compounds decay in water and disinfection by-products production (Clark, 1998; Kastl et al., 1999;Jabari Kohpaei and Sathasivan, 2011; Fisher et al., 2011; Fisher et al., 2017) with advanced hydraulic modelling (Giustolisi et al., 2023) , using an effective and accurate Lagrangian scheme with unlimited pipes parcels.

The novel framework allows analysing the impact of pipe wall biofilm on disinfectant decay, through the definition of two endogenous mechanisms: the reaction of compounds detached by the momentum of water flow and reacting into the bulk while transported downstream, and the local reaction at pipe wall. The proposed framework has been studied with respect to the two distinct effects of the endogenous mechanisms (Case Study 1, small-sized network of Monteparano), and to analyse its support for water quality management through asset management (Case Study 2, large-sized network of Bari).

The discussion of the second case study demonstrates the effectiveness of the proposed framework in directing and exploiting asset management activities to increase water quality.

The novel framework makes water utilities prepared to the use of monitoring data to achieve efficiency of water quality management. The framework can also support the design of the monitoring system and the selection of monitored parameters.

Future research may include the parameters assessment and/or calibration of the second order kinetic model, as well as the calibration of the relevant parameters of the endogenous decay mechanisms, $m_{rc\text{-}wall}$ and $m_{rc\text{-}bulk}$. Also worthy of future research is the analysis of the impact of endogenous mechanisms on private tanks, which are diffuse in mediterranean areas.

## CRediT authorship contribution statement

**D.B. Laucelli:** Formal analysis, Investigation, Methodology, Validation, Visualization, Writing – original draft, Writing – review & editing.

**L. Vergine:** Investigation, Methodology, Validation, Visualization, Writing – original draft, Writing – review & editing.

**G. Messa:** Investigation, Methodology, Validation, Visualization, Writing – original draft, Writing – review & editing.

**O. Giustolisi:** Conceptualization, Data curation, Methodology, Software, Supervision, Writing – review & editing.

## Declaration of Competing Interest

The authors declare that they have no known competing financial interests or personal relationships that could have appeared to influence the work reported in this paper.

## Data Availability

The datasets analysed during the current study, i.e., data of the water distribution network, are not publicly available since they are considered sensitive by water utilities but are available from the corresponding author on reasonable request. All the data generated by analyses during this study are included in this published article.